\documentclass[twocolumn,showpacs,preprintnumbers,amsmath,amssymb,superscriptaddress]{revtex4}
\usepackage{graphicx}% Include figure files
\usepackage{dcolumn}% Align table columns on decimal point
\usepackage{bm}% bold math

\begin{document}

\title{Zigzag graphene nanoribbons without inversion symmetry}% Force line breaks with \\

\author{Lihua Pan}
\affiliation{Complexity Science Center, Yangzhou University, Yangzhou 225002, China}

\author{Jin An}
 \email{anjin@nju.edu.cn}
\affiliation{National Laboratory of Solid State Microstructures and
Department of Physics, Nanjing University, Nanjing 210093, China}

\author{Yong-Jun Liu}
\affiliation{Complexity Science Center, Yangzhou University, Yangzhou 225002, China}

\author{Chang-De Gong}
\email{cdgongsc@nju.edu.cn}
\affiliation{Center for Statistical and Theoretical Condensed Matter
Physics, and Department of Physics, Zhejiang Normal University,
Jinhua 321004, China}

\affiliation{National Laboratory of Solid State Microstructures and
Department of Physics, Nanjing University, Nanjing 210093, China}

\begin{abstract}

Graphene on a substrate will suffer an inversion-symmetry-breaking (ISB) lattice potential. Taking electron-electron interaction into account, we study in this paper the possibility of half-metallicity and noncollinear (NC) magnetic phase for graphene zigzag nanoribbons without inversion symmetry.  At half-filling it is found that half-metallic(HM) state can be achieved at an intermediate value of the ISB potential due to its competition with the electron-electron interaction. Away from half-filling, the phase diagrams of doping versus ISB potential for different ribbon width are given, where the regimes for the HM states and NC magnetic state are clearly indicated and discussed. For ribbons with perfect edges, we predict a topological transition between two HM states with different magnetic structures, which is accompanied by an abrupt transition of electrical conductance along the ribbon from $2e^2/h$ to $e^2/h$.

\end{abstract}

\pacs{72.25.-b, 75.75.-c, 73.22.-f, 71.10.Hf}

\date{\today}% It is always \today, today,

\maketitle

\section{introduction}

Graphene sheet and its related nanostructures have attracted considerable attention owing to their remarkable electronic and structural properties\cite{CastroNeto2009,Geim2007,Abergel2010} and their possible applications in graphene-based spintronics\cite{Novoselov2004,Novoselov2005,Zhang2005}. The free standing graphene crystallite lacks an energy gap and has a Dirac-cone structure due to its negligible spin-orbit interaction. To make graphene be used as a semiconductor, several proposals have been made to open up a tunable gap in its electronic spectra. One is to consider epitaxial graphene on the top of a substrate which breaks the inversion symmetry\cite{Zhou2007,Giovannetti2007,SKim2008,Kwon2009,GLi2009,LKong2010,ZHNi2008,Cuong2011}. An alternative strategy is to construct periodic structures such as antidot lattices\cite{Pedersen2008,Vanevic2009}, graphene ribbons\cite{Berger2006,Han2007}, graphene with regular patterns of hydrogen-covered regions\cite{Sofo2007,Balog2010} or with patterned defects\cite{Silva2010,Appelhans2010}.

The zigzag terminated graphene nanoribbon(ZGNR) has attracted more attentions because it presents a band of zero-energy modes which is the surface states living near the edge of the nanoribbon\cite{Fujita1996,Nakada1996,Kobayashi2005,Niimi2006}. According to Stoner criterion, the ground state has a ferromagnetic instability due to electron-electron interactions and will then lead to an antiferromagnetic(AF) structure where each edge is ferromagnetically polarized but coupled with each other antiferromagnetically\cite{Fujita1996,Hikihara2003}. Most theoretical studies have focused on properties of the neutral ZGNR. Several related works\cite{Jung2009a,Sawada2009,Jung2010} have considered the carrier density away from half-filling and the stable noncollinear(NC) canted magnetic states are predicted in the low doping regime.

Application of a transverse homogeneous electric fields to ZGNR, half-metallicity\cite{Katsnelson2008}can be realized\cite{Son2006,Kan2007}. This means that a small longitudinal source-drain field could be applied to generate fully spin-polarized currents, i.e., electrons with one spin orientation is metallic while electrons with the other orientation is insulating. Some later works\cite{OHod2007,Kan2008,WWu2010,Dutta2009,Li2009,Oeiras2009,OHod2008,Soriano2010,Gundra2011} verified the promising applications for the future spintronics.

In this paper, we present a theoretical study on the possibility of the half-metallicity and NC magnetic states in the ISB graphene nanoribbons. It is found that the ISB potential together with the carrier density affect dramatically the magnetic ground state and the band structure in ZGNR, leading to phase transitions between a series of phases including HM states.

The paper is organized as follows. In Sec. II we give the model Hamiltonian and its mean-field treatment. In Sec. III we discuss the magnetic structures of the ground states. In Sec. IV we investigate and discuss the HM states and correspondingly give the band structures as well as the phase diagram. In Sec. V, we summarize our results.

\section{model}

We assume that the bulk graphene is subjected to a staggered sublattice potential, which breaks the inversion symmetry of graphene. This is the general situation of the graphene on a substrate. The model Hamiltonian can be written as follows,

\begin{eqnarray}
\nonumber H=-t\underset{<i,j>\sigma}{\sum}(c_{i\sigma}^\dag c_{j\sigma}+h.c.)+\underset{i\sigma}{\sum}(V_{i}-\mu)n_{i\sigma}+\\
U\underset{i}{\sum}(n_{i\uparrow}-1/2)(n_{i\downarrow}-1/2)
\end{eqnarray}

Here $t$ is the nearest-neighbor hopping integral, and $U$ the on-site Coulomb repulsion energy. $V_i$ is the staggered sublattice potential with $V_{i}=V(-V)$ on sublattice A(B), and $\mu$ is the chemical potential. Although the substrate potential breaks the inversion symmetry, the Hamiltonian is still invariant under \emph{PI} operation, where \emph{P} is the particle-hole transformation( $\mathrm{c_{i\sigma}}\mathrm{\rightarrow}\mathrm{\eta}\mathrm{c^\dag_{i\sigma}}$, with $\mathrm{\eta=1(-1)}$ if $\mathrm{i}$ belongs to A(B) sublattice) operator and \emph{I} is the space inversion operator, with the inversion center chosen as the center of one of the central hexagons of the ribbon. In the following, energy is measured in unit of $t\approx2.8eV$.

\begin{figure}
\scalebox{0.8}[0.8]{\includegraphics[0,0][284,211]{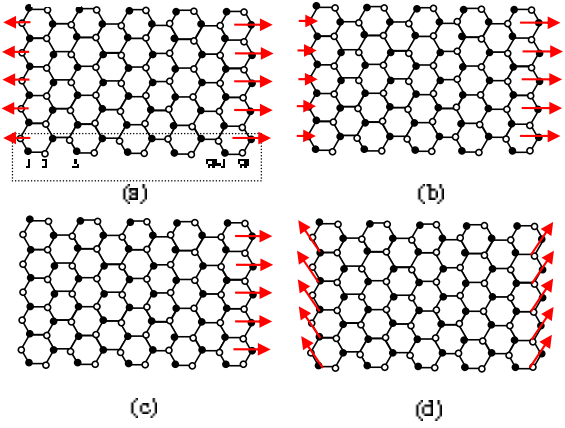}}
\caption{\label{Fig. 1}(Color online) Representative magnetic structures for an ISB ZGNR with width $W=10$. (a) AF ground state for $V=0.10$ and $\delta n=0.1$, where $m_{_\mathrm{L}}=0.091$, $m_{_\mathrm{R}}=0.129$.(b) FM ground state for $V=0.04$ and $\delta n=0.2$, where $m_{_\mathrm{L}}=0.066$, $m_{_\mathrm{R}}=0.135$. (c) Fb ground state for $V=0.10$ and $\delta n=0.4$, where $m_{_\mathrm{L}}=0.0009$, $m_{_\mathrm{R}}=0.125$. (d) NC ground state for $V=0.02$ and $\delta n=0.046$, where $m_{_\mathrm{L}}=0.129$, $m_{_\mathrm{R}}=0.133$ and $\theta=68^{o}$. Here $m_{_\mathrm{L}}$, $m_{_\mathrm{R}}$ are the magnitudes of the spin polarizations on the left and right edges, respectively, whereas $\theta$ is the relative orientation angle between them. The circles(dots) denote the triangular sublattice A(B). The dashed rectangle denotes the unit cell, which is periodically repeated along the vertical direction.}
\end{figure}

To take into account the possibility of the noncollinear spin polarization, the Hubbard term is so decoupled that the mean-field Hamiltonian can be written as,

\begin{eqnarray}
\nonumber \mathcal{H}=-t\underset{\langle i,j\rangle}{\sum}(c_{i}^\dag c_{j}+h.c.)+\\
\underset{i}{\sum}c_{i}^\dag[U(\frac{n_{i}-1}{2}-\mathbf{m}_{i}\cdot\bm{\sigma})+V_{i}-\mu]c_{i}-U\underset{i}{\sum}(\frac{n_{i}^2}{4}-\mathbf{m}_{i}^2)
\end{eqnarray}

where the electron spin polarization vector and the charge density are given by $\mathbf{m}_{i}=(1/2)\langle c_{i}^\dag \bm{\sigma}c_{i}\rangle$, and $n_{i}=\langle c_{i}^\dag c_{i}\rangle$, respectively, with $\bm{\sigma}=(\sigma_{x},\sigma_{y},\sigma_{z})$ the Pauli matrices and $c_{i}^\dag=(c_{i\uparrow}^\dag, c_{i\downarrow}^\dag)$. In the whole paper, we choose $U/t=1.0$, which is an appropriate value consistent with the first-principle calculations \cite{Pisani2007}. For this intermediate $U$, the decoupling process introduced is believed to be reliable and effective. Starting from an initial random spin and density configuration we determine the above parameters for a sample with a strip geometry self-consistently by the standard iteration method.

\begin{figure}
\vspace{-3em}
\scalebox{0.45}[0.45]{\includegraphics[40,40][560,480]{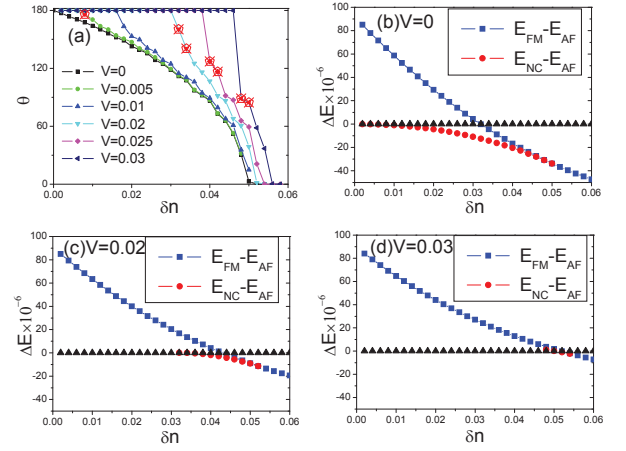}}
\caption{\label{Fig. 2}(Color online)(a): The relative orientation angle $\theta$ between the spin directions on the two edges in the NC canted states as a function of doping at different $V$ for an ISB ZGNR with width $W=10$. The metastable NC states are denoted by the cross marked points. (b)-(d): Energy comparison of the three magnetic states(NC, AF and FM) as a function of doping for the same ribbon at different $V$: (b) $V=0$, (c) $V=0.02$, and (d) $V=0.03$. The energies are all shifted by that of the AF state $E_{_\mathrm{AF}}$ at the corresponding doping value, so as a result the triangles at the zero energy denote the AF state. The filled circles(squares) denote the energy difference between the the NC(FM) and AF states.}
\end{figure}

\section{The magnetic structures}

Similar to ZGNR with the inversion symmetry, the spins at the edges of the sample will be polarized and the solutions can be classified into several categories according to the coupling between the spin polarizations at the two edges. For different ISB potentials and different carrier densities, one can find the following ground state solutions: the AF state and ferromagnetic(FM) state, where the spin orientations at the two edges are antiparallel and parallel to each other respectively; the so called Fb state\cite{Jung2009a}, where only one edge is spin polarized; and the non-collinear(NC) canted state where there is a relative orientation angle $\theta$ between the two spin directions. The corresponding magnetic structures are schematically shown in Fig.~\ref{Fig. 1}, where the effective doping value $\delta n=(1-\langle N\rangle)\times 2W$ with $W$ the width of ribbon and $\langle N\rangle$ the average electron number per site, is measured as the density departure from half-filling per unit cell of ribbon.

The NC canted state is found to be the ground state only in the very low doping level and small $V$. With the increase of $V$ or ribbon width $W$, the doping regime in which NC is favorable will become narrower and finally disappear(see the phase diagram in the next section). In Fig.~\ref{Fig. 2}(a) we plot $\theta$ as a function of doping at different $V$. For a very small ISB potential, $\theta$ gradually decreases from $180^{o}$ to $0^{o}$, whereas for a finite ISB potential, NC solutions can be found only above a critical doping level which also increases with $V$. Some NC solutions are actually metastable and have a higher energy than the AF solutions. In Fig.~\ref{Fig. 2}(b)-(d) we compare the energy of the NC state with that of the AF and FM states, which clearly show the stability of the NC ground state at low-doping levels.

\begin{figure}
\scalebox{0.5}[0.5]{\includegraphics[0,0][448,524]{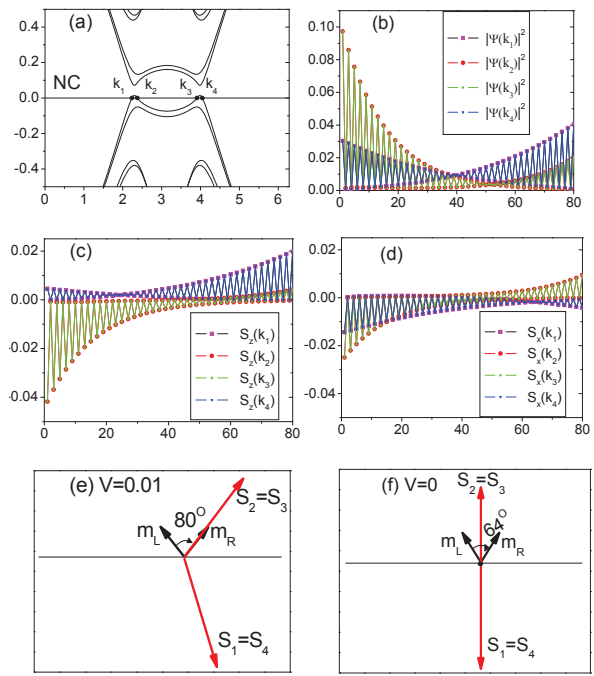}}
\caption{\label{Fig. 3}(Color online)(a) Typical band structure for a NC metallic state at $V=0.01$ and $\delta n=0.024$ for ZGNR with width $W=40$, where the angle $\theta=80^{o}$ and $m_{_\mathrm{R}}\approx m_{_\mathrm{L}}=0.138$. There are four counter-propagating current-carrying states $\mathbf{k_{i}}(i=1,2,3,4)$ at the Fermi level. Correspondingly, the electron density and spin density distributions for the four states are shown in (b)-(d). If the spin axis is so chosen that the edge polarizations can be written as $\mathbf{m_{_\mathrm{R(L)}}}=(\pm sin(\theta/2),0,cos(\theta/2))$, the averaged spin carried by the four states can be expressed as $\mathbf{S_{i}}=m_{i}(cos(\theta_{i}),0,sin(\theta_{i}))$, where $m_{1}=m_{4}=0.400$, $m_{2}=m_{3}=0.353$, and $\theta_{1}=\theta_{4}=-72.3^{o}$, $\theta_{2}=\theta_{3}=51.5^{o}$. As a comparison, the result for the NC canted state with the same doping $\delta n=0.024$ for $V=0$ case is given as follows: $\theta=64^{o}$, $m_{_\mathrm{R}}=m_{_\mathrm{L}}=0.138$, $m_{1}=m_{2}=0.424$, $m_{3}=m_{4}=0.299$, and $\theta_{1}=\theta_{4}=-90^{o}$, $\theta_{2}=\theta_{3}=90^{o}$. These vector quantities are schematically shown in (e) and (f).}
\end{figure}

These NC canted states are found to be always metallic. The representative band structure is shown in Fig.~\ref{Fig. 3}(a), which is characterized by four counter-propagating current-carrying states at the chemical potential, leading to a quantized electric conductance $2e^2/h$ along the ribbon with perfect edges. These four partially polarized current-carrying states(see Fig.~\ref{Fig. 3}(c)-(d)) are not edge ones but extended along the transverse direction even for a wider ribbon(see Fig.~\ref{Fig. 3}(b)), which can be due to the relatively strong coupling between two edges. Detailed calculation shows that the NC canted state for $V=0$ case shares the same picture, but it has a special property which is not preserved by $V\neq 0$ case that the averaged spin carried by the four propagating states is exactly parallel or anti-parallel to the vector sum of the two edge polarizations $\mathbf{m_{_\mathrm{R}}}+\mathbf{m_{_\mathrm{L}}}$(see the caption of Fig.~\ref{Fig. 3}).

\section{Half-metallic states and the phase diagram}
\subsection{Half-filling}
\subsubsection{Classification of the ISB graphene nanoribbons}
\begin{figure}
\scalebox{0.45}[0.45]{\includegraphics[13,30][555,633]{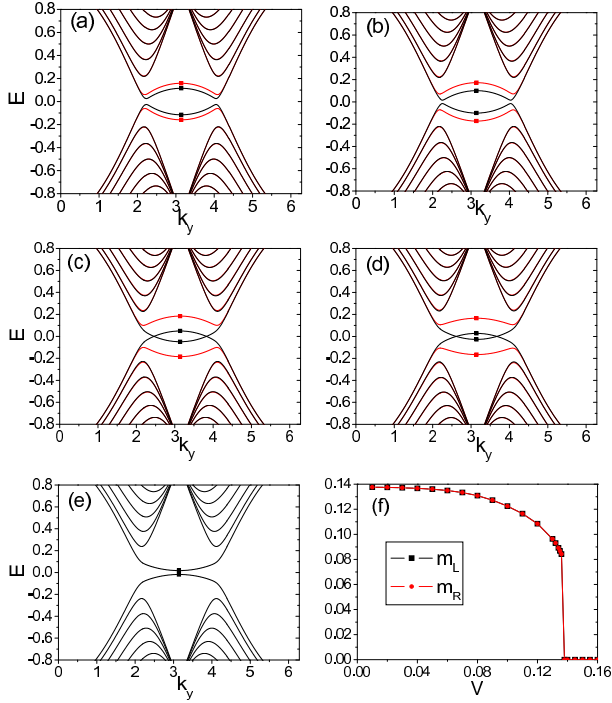}}
\caption{\label{Fig. 4}(Color online)(a)-(e):Band structures of ZGNR with width $W=20$ at half-filling for different ISB potential $V$: (a)
$V=0.03$; (b) $V=0.05$; (c) $V=0.11$; (d) $V=0.13$; (e) $V=0.15$. From (a) to (d), the ground state has the AF magnetic structure and the red and black lines denote the spin-up and spin-down states respectively. The ground state in (e) is a non-magnetic one. The denoted squares are the states at $k_{y}=\pi$ which are completely localized at the zigzag boundary sites. Correspondingly, the spin polarizations at the leftmost and rightmost boundary sites are shown in (f) as functions of $V$.}
\end{figure}

There are two competing processes in the ISB graphene nanoribbons. One is electron-electron interactions, the other is the ISB external field. In a neutral ZGNR with the inversion symmetry($V=0$), the ground state has the AF magnetic structure, which is a consequence of the flat subbands at zero energy. A large magnetic moment emerges around the edge sites even for an infinitesimally small Coulomb $U$, inducing a finite gap\cite{Rossier2008,Jung2009b}as a result. On the other hand, if we neglect electron-electron interaction, application of an ISB external field will mix the wave functions at the two sublattices and will then open up a bulk band gap in neutral ZGNR. Since either of the processes prefers to open up a gap, it is quite unusual that when both of them are present in a graphene ribbon, a state with a negligible gap can be achieved.

\begin{figure}
\scalebox{0.55}[0.55]{\includegraphics[0,0][421,376]{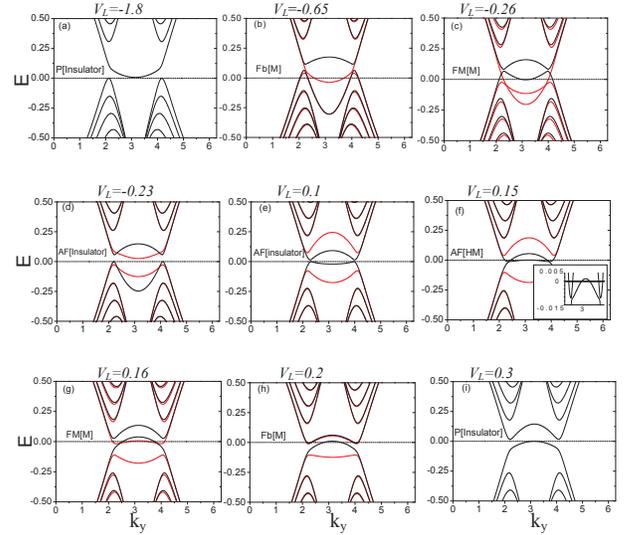}}
\caption{\label{Fig. 5}(Color online)Band structure of the ground states as a function of edge potential for a ZGNR with width $W=20$ and $V=0.05$. Here the right-most site potential $V_{_\mathrm{R}}$ is fixed to be zero, whereas the left-most site potential $V_{_\mathrm{L}}$ is tuned from being negative to positive.}
\end{figure}

For an ISB graphene ribbon, it is found that there exists a ribbon-width dependent threshold value $V_{c}$ of $V$, beyond which the system becomes paramagnetic(PM). When $V<V_{c}$, the ground state at half-filling ($\mu=0$) has an AF magnetic structure which preserves the \emph{PI} symmetry:
\begin{equation}
n_{_\mathrm{R}}+n_{_\mathrm{L}}=2, m_{_\mathrm{R}}=-m_{_\mathrm{L}},
\end{equation}
implying antisymmetric and equal spin polarization on both edges, with $n_{_\mathrm{R(L)}}$ the electron density at the rightmost(leftmost) boundary sites. Therefore, the electronic spectra for the spin-up and spin-down components can always be decoupled and both has the symmetry about zero energy. In Fig.~\ref{Fig. 4}, band structure for a neutral ZGNR with width $W=20$ for both spin directions are shown. In each sub-figure, the center four bands represent that for the edge states. Among the four bands, near $k_{y}=\pi$, the red and black ones below(above) zero energy correspond to spin-up edge states on the right(left) boundary and spin-down ones on the left(right) boundary respectively. Upon increasing $V$, though the bulk gap remains open, the gap between the two spin-down edge states is becoming smaller. When $0.116<V<0.136=V_{c}$, the gap is vanishingly small(less than $10^{-4}t$) while spin-up edge states are still gapful. Thus, a HM state is realized in this regime. We note that the situation is quite similar to that studied in the HM state in ZGNR when applying a transverse electric field\cite{Son2006}, which also breaks the inversion symmetry of graphene ribbon.

The reason of the gap opening and closing can be attributed to the interaction between the left and right spin-down edge states, which is similar to the gap-opening in finite quantum spin Hall systems\cite{Zhou2008}. According to the perturbation theory, the gap due to their interaction can be expressed as $\Delta/t=|\Psi_{_\mathrm{R}}^{*}(k^{*},\mathbf{x}-\mathbf{i})\Psi_{_\mathrm{L}}(k^{*},\mathbf{x})|$. Here the ``crossing point'' $\mathrm{k^{*}}$ is the wavevector where gap is opened up, and $\Psi_{_\mathrm{L}}(k^{*},\mathbf{x})$, $\Psi_{_\mathrm{R}}(k^{*},\mathbf{x})$ are the corresponding spin-down edge-state wave functions for the half-infinite ZGNR with left and right zigzag edges respectively, with \textbf{x} taken to be the leftmost site of the ribbon unit cell and $\mathbf{i}=(1,0)a$. Actually, the states at $k_{y}=\pi$ are completely localized at the zigzag boundary sites. Their eigenenergies can be given analytically by,
\begin{equation}
E_{_\mathrm{L\sigma}}(k_{y}=\pi)=V+U(n_{_\mathrm{L}}-1)/2-\sigma\mathrm{m_{_\mathrm{L}}}
\end{equation}
\begin{equation}
E_{_\mathrm{R\sigma}}(k_{y}=\pi)=-V+U(n_{_\mathrm{R}}-1)/2-\sigma\mathrm{m_{_\mathrm{R}}}
\end{equation}
satisfying $E_{_\mathrm{L\sigma}}(k_{y})=-E_{_\mathrm{R\sigma}}(k_{y})$. So the edge states near $k_{y}=\pi$ are well localized and their decaying lengthes are very small, leading to a vanishingly small overlap and then a vanishingly small gap. On the contrary, the edge states away from $k_{y}=\pi$ are much more spread and have a relatively large overlap so a finite gap is opened up. When $V>V_{c}$, the AF magnetic structure is unstable and the ZGNR will undergo a phase transition from the AF HM state to a charge-density-wave insulator PM state which has a charge gap of magnitude of $2V-U(n_{_\mathrm{R}}-n_{_\mathrm{L}})/2$(see Fig.~\ref{Fig. 4}(e)). The above discussion is based on the result of a graphene ribbon with a fixed width $W=20$, but detailed calculations show that although the threshold $V_{c}$ and the band gap may vary quantitatively a little with the width, the fact of the existence of HM regime is not changed qualitatively.

Generally, graphene on a substrate will inevitably suffer a Rashba spin-orbit (RSO) interaction $H_{_\mathrm{R}}=i\lambda\underset{<i,j>}{\sum}c_{i}^\dag(\bm{\sigma}\times\mathbf{d_{ij}})_{z}c_{j}$, where $\lambda$ is the coupling strength and $\mathbf{d_{ij}}$ is the unit vector along ij direction. The most significant effect of RSO interaction on magnetism is that the spin polarization will be locked to an easy plane which is normal to the graphene plane and zigzag line. For a small $\lambda$, the relative angle between the two edge polarizations will be slightly deviated from $180^{o}$, and the deviation angle $\delta\theta$ is nearly linearly dependent of $\lambda$ with $\delta\theta/\delta\lambda$ estimated as $\sim0.5^{o}/meV$ by our numerical calculation. In actual situation, this interaction is very small and can be neglected in most of applications. Furthermore, though RSO interaction breaks the inversion symmetry, it can be easily checked that this term still preserves the \emph{PI} symmetry of our model Hamiltonian. Therefore, qualitatively RSO interaction should only cause very little deviation from our conclusion on half-metallicity and we then neglect its effect in the following discussion.

\begin{figure}
\scalebox{0.7}[0.7]{\includegraphics[0,0][277,432]{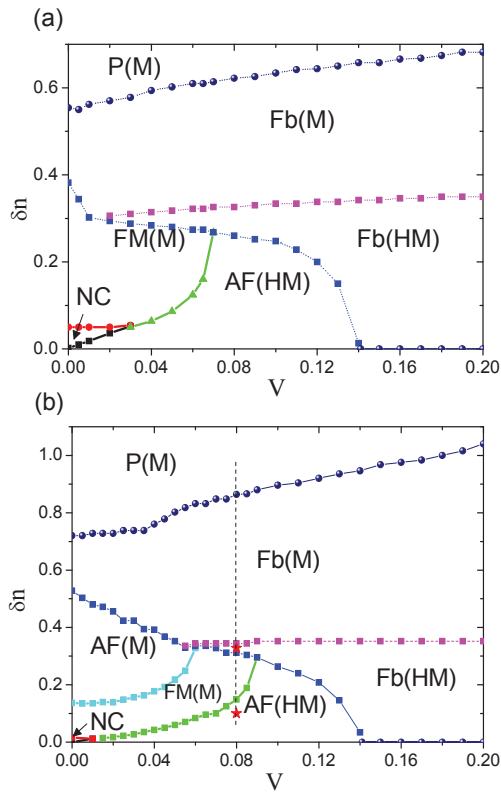}}
\caption{\label{Fig. 6}(Color online)The phase diagrams: doping value $\delta n$ versus the ISB potential $V$ for a ZGNR with $W=10$ for (a) and $W=40$ for (b). The dashed line shows the sequence of phase transitions with increasing doping for a particular fixed $V$.}
\end{figure}

\subsubsection{Manipulating graphene properties by edge potentials }

Since the HM state only exists within a limited small region of $V$ in neutral ZGNR, it is meaningful if one can find some methods to realize half-metallicity in other restricted region. One way to do this is by modifying the edge potentials, since the edge states are localized around the sample boundary, and so they are sensitive to the variation of on-site potentials on the boundary. In a free-electron ZGNR system with broken bulk inversion symmetry, this problem with varying one edge potential has been studied recently in terms of valley Hall effect and it is found that the edge bands can be continuously changed by tuning the on-site edge potentials\cite{Yao2009}.

Here we examine how to manipulate ZGNR to achieve HM state in the presence of electron-electron interactions. Depending on the values of $V_{_\mathrm{L}}$ and $V_{_\mathrm{R}}$, half-filling ZGNR can have various magnetic structures, which affect dramatically the edge bands. In Fig.~\ref{Fig. 5}, for a fixed $V$ we give the band structure of the ground states on a $W=20$ ZGNR as $V_{_\mathrm{L}}$ is tuned from negative to positive value with $V_{_\mathrm{R}}=0$ left unchanged. With variation of $V_{_\mathrm{L}}$, the system undergoes a series of phase transitions due to the magnetic transitions at the edges. When $V_{_\mathrm{L}}$ deviates from zero a little, the ground state still has an AF magnetic structure, but asymmetric(Fig.~\ref{Fig. 5}(d)(e)(f)). In a narrow regime around $V_{_\mathrm{L}}\sim 0.15$, one can achieve an AF HM state(Fig.~\ref{Fig. 5}(f)). In this state, the two spin-down edge bands are mixed due to their interactions. Both of the spin-down edge bands are partially filled with the spin-up edge bands still gapful, leading to the half-metallicity of the state. For a perfect ribbon, this HM state is characterized by a quantized spin-down conductance $3e^2/h$, since there exist one hole pocket and two electron pocket at the Fermi level and thus three pair of propagating modes for the spin-down edge bands(see the inset of Fig.~\ref{Fig. 5}(f)).

\begin{figure}
\scalebox{0.65}[0.65]{\includegraphics[0,0][355,389]{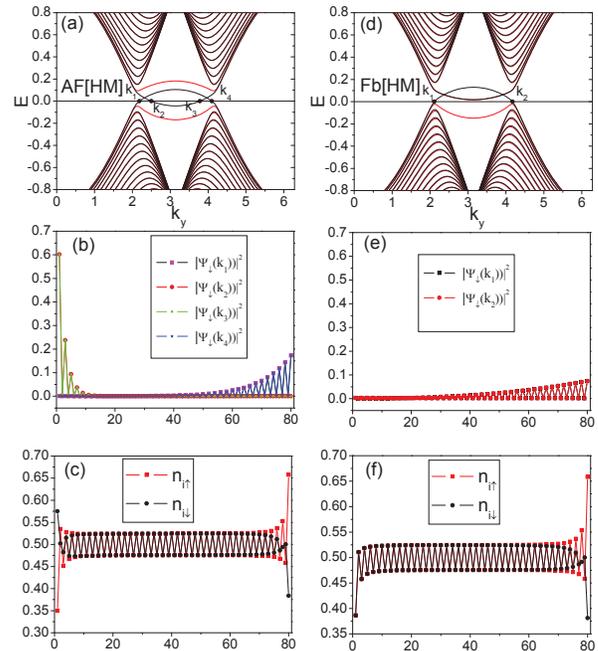}}
\caption{\label{Fig. 7}(Color online)Two HM states for an ISB ZGNR with width $W=40$ and $V=0.08$, indicated by the two star marked points in Fig.~\ref{Fig. 6}. The left column: the HM state with the AF magnetic structure, where $\delta n=0.1$. The right column: the HM state with the Fb structure, where $\delta n=0.328$. Accordingly, the band structures are shown in the top panels, and the density distributions of the edge states at the chemical potentials are shown in the middle panels, whereas the spin density for the ground states are shown in the bottom panels, respectively.}
\end{figure}

\subsection{Away from half-filling}

When the ISB ZGNR is doping away from half-filling, the HM state with the AF magnetic structure can still survive, except breaking the \emph{PI} symmetry spontaneously. This is consistent with the case of doped ZGNR under a transverse electric field, where an AF half-metallic state is found to exist\cite{Jung2010}. To have a concrete picture, we give the phase diagram for ZGNRs in Fig.~\ref{Fig. 6}. Different from the case at half-filling, another type of HM state with the Fb magnetic structure emerges at finite doping and relatively large ISB potential $V$. This state is HM because both the spin-down band are partially filled, similar to that induced by edge potential discussed above. When the ISB potential is within this regime, the system at half-filling is actually a PM insulator, which means that any small doping away from half-filling will induce a magnetic transition and then lead to the phase transition from PM insulator to Fb half-metal. Different to the AF HM state which has four edge states propagating along the two zigzag boundaries, this Fb HM state has only two edge states which are counter-propagating along the polarized right edge(see Fig.~\ref{Fig. 7}). Therefore, this Fb HM state is characterized by a quantized electrical conductance with value $e^2/h$, not $2e^2/h$. The phase diagram for a wider ZGNR(Fig.~\ref{Fig. 6}(b)) is qualitatively the same as that for a narrower one(Fig.~\ref{Fig. 6}(a)), except that there emerges an additional AF metallic state in the former case. When $V=0$, our result is consistent with the previous work\cite{Jung2009a}. We note that in epitaxial graphene a molecular-doping induced metal-insulator transition is observed in ARPES experiment\cite{SYzhou2008}. Moreover, a recent first-principle calculation on epitaxial graphene on SiC has found the evidence of HM state\cite{Huang2011}.

With the increase of $\delta n$, one can realize in order the following phases from AF insulator: AF half-metal, FM metal, Fb half-metal, Fb metal and PM metal(see the dashed line in Fig.~\ref{Fig. 6}(b)). Since graphene on different substrates suffer different ISB potential\cite{Zhou2007,Giovannetti2007,SKim2008,Kwon2009,GLi2009,LKong2010,ZHNi2008,Cuong2011}, upon doping graphene with fixed ISB potential will undergoe phase transitions between HM states and normal-metal states. Particularly, graphene upon doping can change by a first-order transition from one HM state to another, since there exists a phase boundary between the two types of the HM states. This process may be served as a kind of channel switch, which ``close'' the propagating channel on the left edge by spin depolarization upon doping(see Fig.~\ref{Fig. 7}). This is expected to have some important applications in spin related transport.

\section{Summery}
In summery, the magnetic and band structures of an ISB graphene ZGNR have been investigated. At half filling, half-metallicity is found to be realizable for an intermediate ISB potential because of its competition with the electron-electron interactions. For doping away from half-filling, phase diagrams for varying ribbon width are given and another type of HM state is achieved with different magnetic structure. Spin canted states are found to exist only at very low doping levels and low ISB potential. Due to the importance of half-metallicity in spin related transport, these results may have important applications in graphene based spintronics.

\begin{acknowledgments}
This work was supported by NSFC Projects No.10947157
, No.10874073, and 973 Projects No. 2011CB922101.
\end{acknowledgments}


\end{document}